\documentclass[aps,pra,twocolumn,amsmath,showpacs]{revtex4}
\usepackage{bm}
\usepackage{graphicx}
\begin{document}
\setlength{\arraycolsep}{2pt}
\title{Entanglement criteria via the uncertainty relations in SU(2) and SU(1,1) algebra:\\
detection of non-Gaussian entangled states}
\author{Hyunchul Nha$^*$ and Jaewan Kim}
\affiliation{School of Computational Sciences, Korea Institute for Advanced Study, Seoul, Korea} 
\date{\today}
\begin{abstract}
We derive a class of inequalities, from the uncertainty relations of the SU(1,1) and the SU(2) algebra 
in conjunction with partial transposition, that must be satisfied by any separable two-mode states. 
These inequalities are presented in terms of the SU(2) operators 
$J_x=\left(a^\dag b+ab^\dag\right)/2$, $J_y=\left(a^\dag b-ab^\dag\right)/2i$, 
and the total photon number $\langle N_a+N_b\rangle$. 
They include as special cases the inequality derived by Hillery and Zubairy [Phys. Rev. Lett. {\bf 96}, 050503 (2006)], 
and the one by Agarwal and Biswas [New J.~Phys.~{\bf 7}, 211 (2005)]. 
In particular, optimization over the whole inequalities leads to the criterion 
obtained by Agarwal and Biswas. 
We show that this optimal criterion can detect entanglement 
for a broad class of non-Gaussian entangled states, i.e., the SU(2) minimum-uncertainty states. 
Experimental schemes to test the optimal criterion are also discussed, 
especially the one using linear optical devices and photo detectors.
\end{abstract}
\maketitle

\narrowtext
\section{Introduction}
Entanglement, one of the defining properties of quantum mechanics, 
is a key element for quantum information processing using discrete or continuous variables (CVs)\cite{QIbook}. 
In many applications, it is of great importance to verify entanglement for multipartite systems in one way or another. 
One possible approach is to derive the ``classical" limit attainable by classical means 
for a specific quantum protocol. Experimental demonstration of surpassing this limit 
can be an indirect proof of entanglement. 
For example, in the CV quantum teleportation of coherent states, 
the fidelity larger than 1/2 may indicate the presence of entanglement 
in the state shared by two parties\cite{Braunstein}. 
Another approach is to derive the inequalities that all separable states must satisfy. 
Violation of such inequalities is sufficient, though not necessary in general, for demonstrating entanglement. 
Previously, those inequalities were derived for CVs in terms of the variances of the canonical operators 
${\hat x}_i$ and ${\hat p}_i$ ($i=1,2$), 
or the quadrature amplitudes for optical fields\cite{Duan,Simon,Mancini,Eisert}. 
In particular, those conditions were proved to be both sufficient and necessary to manifest entanglement 
for bipartite Gaussian states\cite{Duan,Simon}.

Non-Gaussian states, however, are also important and even essential in some cases\cite{nha}. 
It is therefore crucial to have entanglement criteria applicable beyond Gaussian states 
for further applications. 
Working in this direction, several authors recently obtained some inequalities 
by considering the SU(2) and the SU(1,1) algebra\cite{Agarwal,Hillery1}. 
In particular, Agarwal and Biswas used the negativity of partial transposition (NPT) 
to derive a separability condition for testing single-photon entangled states, 
$\alpha|1,0\rangle+\beta|0,1\rangle$\cite{Agarwal}. 
Interestingly, Shchukin and Vogel showed that all the previously known criteria for CVs 
in Refs.~\cite{Duan,Simon,Mancini} and in Refs.~\cite{Agarwal,Hillery1} can be derived in principle 
by taking into account the NPT condition for a hierachy of two-mode moments\cite{Shchukin1,Peres1}. 
In practice, however, measurement of higher order moments needed particularly 
in the approach of Ref.~\cite{Shchukin1} seems to be rather demanding\cite{Shchukin2}. 
For practical applications, it may be desirable to have inseparability criteria that can be tested 
in experiment with the least possible resources\cite{Barnum}. 
In addition, once a certain criterion is derived, it is necessary to identify the class of states that can be 
detected by such a criterion.

In this paper, we study in some detail the separability conditions that can be obtained 
from the uncertainty relations in the SU(2) and the SU(1,1) algebra in conjunction with partial transposition. 
The SU(2) algebra deals with the angular momentum operators $J_x,J_y$ and $J_z$, 
which obey the commutation relations $\left[J_i,J_j\right]=i\epsilon_{ijk}J_k$ $(i,j,k=x,y,z)$. 
This algebra can be realized in optics using two mode fields represented by the annihilation operators $a$ and $b$, 
as 
\begin{eqnarray}
J_x&=&\frac{1}{2}\left(a^\dag b+ab^\dag\right),\nonumber\\ 
J_y&=&\frac{1}{2i}\left(a^\dag b-ab^\dag\right),\nonumber\\ 
J_z&=&\frac{1}{2}\left(a^\dag a-b^\dag b\right).
\label{eqn:su2operators}
\end{eqnarray}
On the other hand, the SU(1,1) algebra with the operators $K_x,K_y$ and $K_z$ that satisfy 
$\left[K_x,K_y\right]=-iK_z,\left[K_y,K_z\right]=iK_x$, and $\left[K_z,K_x\right]=iK_y$, 
can be realized as
\begin{eqnarray}
K_x&=&\frac{1}{2}\left(a^\dag b^\dag+ab\right),\nonumber\\ 
K_y&=&\frac{1}{2i}\left(a^\dag b^\dag-ab\right),\nonumber\\ 
K_z&=&\frac{1}{2}\left(a^\dag a+b^\dag b+1\right).
\label{eqn:su11operators}
\end{eqnarray}
Starting from the uncertainty relations in the SU(1,1) algebra 
and applying the partial transposition, 
we will derive a class of inequalities, in a sum form of the variances of the SU(2) operators,  
that must be satisfied by all separable states. 
These include as special cases the inequality derived by Hillery and Zubairy\cite{Hillery1}, 
and the one by Agarwal and Biswas\cite{Agarwal}. 
In particular, the inequality optimized over the whole inequalities 
is none other than the one obtained by Agarwal and Biswas\cite{Agarwal}. 
Importantly, it will be clarified that this optimal criterion can be implemented in experiment 
by measuring the variances $\Delta J_x$, $\Delta J_y,$ 
and the mean photon number $\langle a^\dag a+b^\dag b\rangle$, 
as is the case with the inequality by Hillery and Zubairy\cite{Hillery1}. 
Furthermore, we will show that the optimal inequality can detect entanglement for a broad class of 
non-Gaussian entangled states, i.e., the SU(2) minimum-uncertainty states. 

The minimum-uncertainty states for the SU(2) algebra were first derived 
by C.~Aragone {\it et al.}\cite{Aragone}, 
and have long attracted much of theoretical interest\cite{Wodkiewicz,Agarwal2,Luis,Gerry,Yurke,Hillery2}.
On an application side, the SU(2) squeezed states have been proposed 
to improve the accuracy of phase measurement in quantum interferometer\cite{Yurke,Hillery2}.
Furthermore, it may be natural to expect more useful applications because the SU(2) minimum-uncertainty states 
are non-Gaussian entangled ones. 
For example, it was recently shown that the Gedanken Bell experiment involving the spin singlet states 
proposed by A.~Peres\cite{Peres2} can be approximately realized 
in quantum optics using the SU(2) coherent states\cite{Gerry}. 
These states belong to a subclass of the SU(2) minimum-uncertainty states under study in this paper.  
Our approach thus suggests in a sense an alternative method to verify entanglement in those states. 

This paper is organized as follows. 
In Sec.~II we briefly introduce the partial transposition within the phase-space description.
In Sec.~III, we derive a class of inequalities in an arbitrary sum form by combining the uncertainty relation 
between $K_x$ and $K_y$ of the SU(1,1) algebra with the notion of partial transposition. 
In particular, the strongest inequality among them is obtained that is expressed in terms of the SU(2) operators 
$J_x$, $J_y$, and the total photon number $N_a+N_b$. 
In Sec.~IV, we present the SU(2) minimum-uncertainty states and 
show that the strongest criterion can detect entanglement for all those states. 
In Sec.~V, experimental schemes to generate the minimum-uncertainty states and those to measure the observables 
$J_x$ and $J_y$, especially a linear optical scheme, are discussed. 
Finally, we summarize our results in Sec.~VI. 

\section{Partial transposition}
In this section, we briefly introduce partial transposition of a density operator in the phase space. 
Let us first consider a single mode field in the position representation, 
$\rho=\int dxdx'\rho_{xx'}|x\rangle\langle x'|$,
where $\hat x|x\rangle=x|x\rangle$. 
The position $\hat x$ and the momentum $\hat p$ are defined via the relations $a=(\hat x+i\hat p)/\sqrt{2}$, 
and $a^\dag=(\hat x-i\hat p)/\sqrt{2}$. 
The characteristic function $C_{\rho^T}(\lambda)\equiv{\rm Tr} \{\rho^TD(\lambda)\}$ 
for the transposed density operator $\rho^T=\int dxdx'\rho_{xx'}|x'\rangle\langle x|$ is then given by 
\begin{eqnarray}
C_{\rho^T}(\lambda)
&=&\int dxdx'\rho_{xx'}\langle x|D(\lambda)|x'\rangle\nonumber\\
&=&\int dxdx'\rho_{xx'}\langle x'|D(-\lambda^*)|x\rangle\nonumber\\
&=&C_{\rho}(-\lambda^*),
\end{eqnarray}
where $C_{\rho}(\lambda)$ is the characteristic function of the original state $\rho$, 
and $D(\lambda)=e^{\lambda a^\dag-\lambda^*a}$ is the displacement operator.
Therefore, the $s$-ordered distribution 
\cite{Barnett} of $\rho^T$ is related to that of $\rho$, as
\begin{eqnarray}
W_{\rho^T}(\alpha,s)&=&\frac{1}{\pi^2}\int d^2\lambda 
e^{\alpha\lambda^*-\alpha^*\lambda}e^{s|\lambda|^2/2}C_{\rho^T}(\lambda)\nonumber\\
&=&\frac{1}{\pi^2}\int d^2\lambda 
e^{\alpha\lambda^*-\alpha^*\lambda}e^{s|\lambda|^2/2}C_{\rho}(-\lambda^*)\nonumber\\
&=&W_{\rho}(\alpha^*,s).
\end{eqnarray}
That is, $W_{\rho^T}(\alpha_x,\alpha_y,s)=W_{\rho}(\alpha_x,-\alpha_y,s)$ for all $s$ ($-1\le s\le1$). 
This reflects the fact that transposition physically represents the motion-reversed state\cite{Peres1,Simon}.

From now on, let us work in the Glauber-P representation $(s=1)$. 
A normally-ordered moment $\langle a^{\dag m}a^n\rangle_{\rho^T}$ of the transposed density operator is 
expressed in terms of another moment of the original state, as 
\begin{eqnarray}
\langle a^{\dag m}a^n\rangle_{\rho^T}&=&\int d^2\alpha \alpha^{*m}\alpha^n
P_{\rho^T}(\alpha_x,\alpha_y)\nonumber\\
&=&\int d^2\alpha \alpha^{*m}\alpha^n P_{\rho}(\alpha_x,-\alpha_y)\nonumber\\ 
&=&\int d^2\alpha \alpha^{m}\alpha^{*n} P_{\rho}(\alpha_x,\alpha_y)\nonumber\\
&=&\langle a^{\dag n}a^m\rangle_{\rho}.
\end{eqnarray}
For an arbitrary operator $\hat O$ represented in the normal ordering as $\hat O=\Sigma C_{mn}a^{\dag m}a^n$,
we obtain $\langle \hat O\rangle_{\rho^T}=\Sigma C_{mn}\langle a^{\dag n}a^m\rangle_{\rho}$. 
If the coefficients $C_{mn}$ are all real, 
we simply have $\langle \hat O\rangle_{\rho^T}=\langle {\hat O}^\dag\rangle_{\rho}$.

Extension of the previous results to the partial transposition for the multi-mode case is straightforward. 
For example, in the case of partial transposition for mode $b$,  
we have 
\begin{eqnarray}
\langle a^{\dag m}a^nb^{\dag p}b^q\rangle_{\rho^{PT}}=\langle a^{\dag m}a^nb^{\dag q}b^p\rangle_{\rho}.
\label{eqn:pt}
\end{eqnarray}

\section{Separability conditions}
In this section, we will derive a class of inequalities via uncertainty relations among the SU(1,1) operators 
along with the partial transposition considered in Sec.~II. 
To begin with, from the commutator $\left[K_x,K_y\right]=-iK_z$, we have the uncertainty relation 
$\Delta K_x \Delta K_y \ge \frac{1}{2}|K_z|$.
If a two-mode state is separable, it remains physical after partial transposition\cite{Peres1}. 
Thus, the above inequality must be satisfied also by the partially transposed density operator $\rho^{PT}$,
i.e., 
\begin{eqnarray}
(\Delta K_x)_{\rho^{PT}} (\Delta K_y)_{\rho^{PT}} \ge \frac{1}{2}\langle K_z\rangle_{\rho^{PT}},
\label{eqn:uncertainty1}
\end{eqnarray}
if the state is to be separable\cite{Agarwal}. Using Eq.~(\ref{eqn:pt}), 
we find that $(\Delta K_x)_{\rho^{PT}}$ is related to $(\Delta J_x)_{\rho}$, as
\begin{eqnarray}
&&(\Delta K_x)_{\rho^{PT}}^2\equiv\langle K_x^2\rangle_{\rho^{PT}}-\langle K_x\rangle_{\rho^{PT}}^2\nonumber\\
&=&\frac{1}{4}\langle a^{\dag2}b^{\dag2}+a^2b^2+a^\dag ab^\dag b+aa^\dag bb^\dag\rangle_{\rho^{PT}}
-\frac{1}{4}\langle a^{\dag}b^{\dag}+ab\rangle_{\rho^{PT}}^2\nonumber\\
&=&\frac{1}{4}\langle a^{\dag2}b^2+a^2b^{\dag2}+a^\dag ab^\dag b+aa^\dag bb^\dag\rangle_{\rho}
-\frac{1}{4}\langle a^{\dag}b+ab^\dag\rangle^2_{\rho}\nonumber\\
&=&\langle J_x^2\rangle_{\rho}
+\frac{1}{4}-\langle J_x\rangle_{\rho}
^2=(\Delta J_x)_{\rho}^2+\frac{1}{4}.
\label{eqn:uncertainty2}
\end{eqnarray}
Similarly, we obtain 
\begin{eqnarray}
(\Delta K_y)_{\rho^{PT}}^2=(\Delta J_y)_{\rho}^2+\frac{1}{4},
\label{eqn:uncertainty3}
\end{eqnarray}
and 
\begin{eqnarray}
\langle K_z\rangle_{\rho^{PT}}=\langle K_z\rangle_{\rho}.
\label{eqn:kz}
\end{eqnarray}
Now let us construct an arbitrary sum of $(\Delta K_x)_{\rho^{PT}}^2$ and $(\Delta K_y)_{\rho^{PT}}^2$. 
Using $x^2+y^2\ge 2xy$, with $x=\sqrt{\alpha}(\Delta K_x)_{\rho^{PT}}$ and $y=\sqrt{\beta}(\Delta K_y)_{\rho^{PT}}$, 
we have 
\begin{eqnarray}
\alpha(\Delta K_x)_{\rho^{PT}}^2+\beta(\Delta K_y)_{\rho^{PT}}^2\ge \sqrt{\alpha\beta}\langle K_z\rangle_{\rho^{PT}},
\label{eqn:uncertainty4}
\end{eqnarray}
where the inequality (\ref{eqn:uncertainty1}) was used. 
In terms of the moments of the original {\it separable} state $\rho_S$, using the relations 
in Eqs.~(\ref{eqn:uncertainty2}),(\ref{eqn:uncertainty3}), and (\ref{eqn:kz}), 
the inequality (\ref{eqn:uncertainty4}) becomes 
\begin{eqnarray}
(\Delta J_x)_{\rho_S}^2+c^2(\Delta J_y)_{\rho_S}^2\ge 
\frac{1}{2}c\langle N_+\rangle_{\rho_S}-\frac{1}{4}(c-1)^2,
\label{eqn:uncertainty5}
\end{eqnarray}
where $N_+\equiv N_a+N_b$ is the total photon number 
and $c\equiv\sqrt{\alpha/\beta}>0$ is an arbitrary parameter.
On the other hand, for a general state, entangled or not, 
the SU(2) commutator $\left[J_x,J_y\right]=iJ_z$ sets the uncertainty relation 
\begin{eqnarray}
(\Delta J_x)^2+c^2(\Delta J_y)^2\ge 
c|\langle J_z\rangle|=\frac{1}{2}c|\langle N_-\rangle|,
\label{eqn:uncertainty6}
\end{eqnarray}
where $N_-\equiv N_a-N_b$ is the photon number difference. 
Note that the terms on the left-hand sides of Eq.~(\ref{eqn:uncertainty5}) and Eq.~(\ref{eqn:uncertainty6}) are the same.
Therefore, if the quantity on the right-hand side of Eq.~(\ref{eqn:uncertainty5}) is larger than 
that of Eq.~(\ref{eqn:uncertainty6}), 
there can be some entangled states that satisfy the inequality~(\ref{eqn:uncertainty6}), 
but that violate the one in~(\ref{eqn:uncertainty5}). 
The inequalities~(\ref{eqn:uncertainty5}) thereby define a class of separability criteria for CVs.
This is the case when $c$ is chosen in the interval $(c_-,c_+)$, 
where $c_\pm=(1+2N_m)\pm\sqrt{(1+2N_m)^2-1}$ and 
$N_m\equiv{\rm min}\{\langle N_a\rangle,\langle N_b\rangle\}$.

Note that a special case for $c=1$ in~(\ref{eqn:uncertainty5}) is the inequality obtained 
by Hillery and Zubairy in Ref.~\cite{Hillery1}, 
where they used a different procedure to derive the separability condition 
\begin{eqnarray}
(\Delta J_x)^2+(\Delta J_y)^2\ge\frac{1}{2}\langle N_+\rangle. 
\label{eqn:Hillery}
\end{eqnarray}

We now want to optimize the inequality~(\ref{eqn:uncertainty5}) 
by choosing a proper parameter $c=c_{\rm opt}$. This can be done in the following way\cite{Giovannetti}. 
Since our goal is to make the inequality violated by a given entangled state, 
we need to minimize, possibly lower than zero, the following value 
\begin{eqnarray}
&&(\Delta J_x)^2+c^2(\Delta J_y)^2- 
\frac{1}{2}c\langle N_+\rangle+\frac{1}{4}(c-1)^2\nonumber\\
& &=\frac{1}{4}\left[\left(1+4(\Delta J_y)^2\right)\left(c-\frac{1+\langle N_+\rangle}{1+4(\Delta J_y)^2}\right)^2\right.
\nonumber\\
& &\left.\hspace{0.8cm}+1+4(\Delta J_x)^2-\frac{(1+\langle N_+\rangle)^2}{1+4(\Delta J_y)^2}\right].
\label{eqn:min}
\end{eqnarray}
Since $\Delta J_x,\Delta J_y$ and $\langle N_+\rangle$ 
are just real numbers for a given state, 
we can always choose $c=c_{\rm opt}=(1+\langle N_+\rangle)/(1+4(\Delta J_y)^2)>0$. 
Then, the inequality~(\ref{eqn:uncertainty5}) is reduced to 
\begin{eqnarray}
(1+4(\Delta J_x)^2)(1+4(\Delta J_y)^2)\ge(1+\langle N_+\rangle)^2.
\label{eqn:opt}
\end{eqnarray}
This inequality is in fact the one that can be directly obtained from Eq.~(\ref{eqn:uncertainty1}) by inserting the relations 
in Eqs.~(\ref{eqn:uncertainty2}),(\ref{eqn:uncertainty3}), and (\ref{eqn:kz}). 
The inequality~(\ref{eqn:opt}) provides the strongest condition for separability 
among the ones in Eq.~(\ref{eqn:uncertainty5}). 
Indeed, it can be readily checked that in general, regardless of the operator algebra involved, 
the separability condition in the product form, like Eq.~(\ref{eqn:uncertainty1}), 
always gives the optimal inequality among those in a sum form, 
on account of the simple algebraic relation $x^2+y^2\ge 2xy$ used in the derivation\cite{Eisert}. 

The inequality~(\ref{eqn:opt}) was previously derived by Agarwal and Biswas in Ref.~\cite{Agarwal}, 
though not explicitly presented in terms of $\Delta J_x$ and $\Delta J_y$. 
Identification of the terms $\Delta J_x$ and $\Delta J_y$, however, 
becomes important in an experimental point of view, 
and we will discuss it further, particularly their measurement, in Sec.~V.

\section{SU(2) minimum-uncertainty states}

In this section, we apply the criterion~(\ref{eqn:opt}) to detect entanglement 
for the SU(2) minimum-uncertainty states. 
For comparison, however, we also consider the inequality~(\ref{eqn:Hillery}) and investigate its power of detecting inseparability. The SU(2) minimum-uncertainty states were derived 
in the literature\cite{Aragone,Hillery2}, and
we first present those states in the following. 
Then, we will show that the optimal inequality~(\ref{eqn:opt}) can detect entanglement 
for any arbitrary SU(2) minimum-uncertainty states, 
whereas the inequality~(\ref{eqn:Hillery}) is limited in its applicability.

\subsection{Minimum-uncertainty states}

For two general Hermitian operators $u$ and $v$, we have the uncertainty relation 
$\Delta u\Delta v\ge \left|\langle[u,v]\rangle\right|/2$. 
The minimum uncertainty states refer to those which satisfy the equality, i.e., 
$\Delta u\Delta v=\left|\langle[u,v]\rangle\right|/2$. 
These states can be derived by solving the eigenvalue equation\cite{Schiff}
\begin{eqnarray}
(u+i\lambda v) |\Psi\rangle=\beta|\Psi\rangle,
\label{eqn:eigen}
\end{eqnarray}
where $\beta$ is a complex eigenvalue.
After some algebra, it is found that 
\begin{eqnarray}
(\Delta u)^2&=&\frac{1}{2}\left|\lambda\langle[u,v]\rangle\right|,\nonumber\\
(\Delta v)^2&=&\frac{1}{2}\left|\frac{1}{\lambda}\langle[u,v]\rangle\right|.
\label{eqn:eigen-rel6}
\end{eqnarray}
Therefore, $\Delta u\Delta v=\left|\langle[u,v]\rangle\right|/2$ is satisfied, 
and the parameter $|\lambda|$ can be interpreted as the degree of squeezing. 
For $|\lambda|=1$, the two variances are the same, i.e. no squeezing, 
and for $|\lambda|\ne1$, the fluctuation of one variable is reduced 
at the expense of that of the other variable. 

In our case, $u=J_x$ and $v=J_y$, and Eq.~(\ref{eqn:eigen}) becomes
\begin{eqnarray}
\left[(1-\lambda)ab^\dag+(1+\lambda)a^\dag b)\right]|\Psi\rangle_\lambda=2\beta|\Psi\rangle_\lambda. 
\label{eqn:eigen-new}
\end{eqnarray}
We will consider only the case that the mean number of mode $a$ is larger than that of mode $b$, 
$\langle N_a\rangle>\langle N_b\rangle$ ($\lambda>0$), 
because the opposite case can be treated just by interchanging the modes $a$ and $b$. 
Thus, for the SU(2) minimum-uncertainty states, the following relations are satisfied 
\begin{eqnarray}
(\Delta J_x)^2=\frac{\lambda}{4}\langle N_-\rangle,\hspace{1cm}
(\Delta J_y)^2=\frac{1}{4\lambda}\langle N_-\rangle,
\label{eqn:eigen-rel7}
\end{eqnarray}
where $\langle N_-\rangle\equiv\langle N_a\rangle-\langle N_b\rangle$. 
We see that only the calculations of $\langle N_\pm\rangle$ are necessary 
to use the separability conditions~(\ref{eqn:Hillery}) and~(\ref{eqn:opt}), 
due to the relations in Eq.~(\ref{eqn:eigen-rel7}) for nonzero $\lambda$. 

In the following, we will consider the minimum-uncertainty states only for $0\le\lambda\le1$, 
i.e., $J_x$-squeezed states in Eq.~(\ref{eqn:eigen-rel7}). The case for $\lambda>1$, i.e., 
$J_y$-squeezed states, can be treated without further calculation by the following observation; 
If the annihilation operator $b$ is redefined as $be^{-i\frac{\pi}{2}}$, 
the operator $J_x$ is changed to $J_y$ and vice versa. (See Eq.~(\ref{eqn:su2operators}).)  
This implies that local phase-shift of the mode $b$ by the amount of $\frac{\pi}{2}$ 
transforms $J_x$-squeezed states to $J_y$-squeezed ones. 
Moreover, note that the inequalities~(\ref{eqn:Hillery}) and~(\ref{eqn:opt}) 
is symmetric with respect to $J_x$ and $J_y$. 
Therefore, the problem of detecting inseparability for $J_y$-squeezed states is redundant. 

More precisely, using the relation $e^{i\phi b^\dag b}be^{-i\phi b^\dag b}=be^{-i\phi}$, 
the eigenstate $|\Psi\rangle_\lambda$ for $\lambda>1$ ($J_y$-squeezed state) 
is obtained from the state $|\Phi\rangle_{1/\lambda}$ ($J_x$-squeezed state), as 
$|\Psi\rangle_\lambda=e^{-i\frac{\pi}{2}b^\dag b}|\Phi\rangle_{1/\lambda}$, 
which can be easily checked in Eq.~(\ref{eqn:eigen-new}). 

\subsection{SU(2) minimum-uncertainty states}
We now solve Eq.~(\ref{eqn:eigen-new}) to obtain SU(2) minimum-uncertainty states.\\\\
(i) $\lambda=1$;\\\\
In the {\it unsqueezed} case, Eq.~(\ref{eqn:eigen-new}) becomes $a^\dag b|\Psi\rangle=\beta|\Psi\rangle$. 
Using the photon-number basis, $|\Psi\rangle=\Sigma C_{n_1,n_2}|n_1\rangle_a|n_2\rangle_b$, 
we obtain the recurrence relation $C_{n_1-1,n_2+1}\sqrt{n_1(n_2+1)}=\beta C_{n_1,n_2}$ for $n_1\ge1$ and $n_2\ge0$. 
If $\beta$ is nonzero, we additionally have $C_{0,n_2}=0$ for all $n_2\ge0$, 
since $a^\dag b|\Psi\rangle$ does not contain the vacuum-state component for mode $a$. 
We then find no solution, because all the other coefficients also vanish from the recurrence relations. 
The eigenevalue $\beta$ therefore must be zero. 
The same reasoning will be used below for the case of $\lambda<1$.

For $\beta=0$, the recurrence relation suggests that the mode $b$ is in the vacuum state. 
That is, $|\Psi\rangle=|\Phi\rangle_a|0\rangle_b$, where $|\Phi\rangle_a$ is an arbitrary state, 
which represents a product state. Thus, these {\it unsqueezed} states are of no interest in our work\cite{nnha}.
\\\\ 
(ii) $\lambda<1$;\\\\
In the squeezed case, let us introduce a transformed state $|\Psi'\rangle=S^{-1}(z)|\Psi\rangle$, 
where $S(z)\equiv e^{zab^\dag-z^*a^\dag b}$ describes the beam-splitter action\cite{Campos}. 
We choose the complex value $z\equiv re^{i\phi_z}$ to make simple the recurrence relations 
for the state $|\Psi'\rangle$\cite{Hillery2}. 
By inserting $|\Psi\rangle=S(z)|\Psi'\rangle$ to Eq.~(\ref{eqn:eigen-new}), and 
using $S(z)aS^\dag(z)=a\cos{r}+be^{-i\phi_z}\sin{r}$ and $S(z)bS^\dag(z)=b\cos{r}-ae^{i\phi_z}\sin{r}$, 
we get 
\begin{eqnarray}
\left[\sqrt{1-\lambda^2}(N_a-N_b)+2\lambda a^\dag b\right]|\Psi'\rangle=2\beta|\Psi'\rangle,
\label{eqn:eigen-newnew}
\end{eqnarray}
with the choice $\tan r=\sqrt{\frac{1-\lambda}{1+\lambda}}$ and $\phi_z=0$. 

A further simplification is made by the observation that 
the operators in Eq.~(\ref{eqn:eigen-new}) or (\ref{eqn:eigen-newnew}) preserve total photon-number, i.e., 
$N=N_a+N_b$ becomes a constant of motion. 
We can thus construct $N$-manifolds 
such that the eigenstates $|\Psi'\rangle$ for a given $N$ are of the form  
$|\Psi'\rangle=\Sigma_{p}C_p|p\rangle_a|N-p\rangle_b$. 
We then obtain the recurrence relation 
\begin{eqnarray}
2\lambda\sqrt{p(N-p+1)}C_{p-1}
=\beta_{p}C_{p},
\label{eqn:recurrence}
\end{eqnarray} 
where  $\beta_{p}\equiv2\beta-\sqrt{1-\lambda^2}(2p-N)$. 
Clearly, the parameter $\beta_{p}$ must be zero for a certain $p$, otherwise no solution exists. 
( See the argument in (i) for $\lambda=1$.) 
If $\beta_{p}$ vanishes for $p=m$, where $m$ is a non-negative integer ($0\le m\le N$), 
the eigenvalue $2\beta$ is given by $2\beta=\sqrt{1-\lambda^2}(2m-N)$, 
and $|\Psi'\rangle$ contains only the terms for $p\ge m$. 
The parameter $m$ is thereby a truncation number that characterizes distinct eigenstates.
We now obtain $\beta_{p}=\beta_{p}^{(m)}=-2\sqrt{1-\lambda^2} (p-m)$. 

The recurrence relation in Eq.~(\ref{eqn:recurrence}) can be readily iterated to give 
\begin{eqnarray}
|\Psi'\rangle_{N,m}=\frac{1}{W_{Nm}}\Sigma_{p\ge m}C_p^{N,m}|p\rangle_a|N-p\rangle_b,
\label{eqn:eigenstate}
\end{eqnarray} 
where 
\begin{eqnarray}
C_p^{N,m}=\frac{\Lambda^{p-m}}{(p-m)!}
\sqrt{\frac{p!(N-m)!}{m!(N-p)!}}.
\label{eqn:coeff}
\end{eqnarray}
and $\Lambda=-\frac{\lambda}{\sqrt{1-\lambda^2}}$.
The normalization constant $W_{Nm}$ is given by $W_{Nm}={_2F_1}\left[m+1,m-N,1,-\Lambda^2\right]$, 
where $_2F_1$ is the hypergeometric function.
The subscripts/superscripts $\{N,m\}$ are used to classify different eigenstates, 
where $N$ is the total photon number and $m$ is the truncation number in Eq.~(\ref{eqn:eigenstate})\cite{nha1}. 
Note that the states in Eq.~(\ref{eqn:eigenstate}) are generally non-Gaussian entangled states, and in particular, 
the state for $m=0$ describes a two-mode binomial state\cite{Stoler}. 
 
We can calculate $\langle N_-\rangle$ using the transformed state $|\Psi'\rangle$ with the relation 
$S^\dag(z)N_-S(z)=\lambda N_--2\sqrt{1-\lambda^2}J_x$ as 
\begin{eqnarray}
\langle N_-\rangle
=\lambda \langle N_-\rangle_{|\Psi'\rangle}-2\sqrt{1-\lambda^2}\langle J_x\rangle_{|\Psi'\rangle}.
\label{eqn:N-diff}
\end{eqnarray}
Then, the variances $(\Delta J_x)^2$ and $(\Delta J_y)^2$ are given by the relations in Eq.~(\ref{eqn:eigen-rel7}).
Let us first consider some simple SU(2) minimum-uncertainty states and see whether entanglement in those states 
can be detected using the inequality~(\ref{eqn:Hillery}) or~(\ref{eqn:opt}). \\

(a) case of $m=N$\\
The eigenstates are simply $|\Psi'\rangle_{N,N}=|N\rangle_a|0\rangle_b$ in Eq.~(\ref{eqn:eigenstate}), 
and, in the original frame, they become  $|\Psi\rangle_{N,N}=S(z)|N\rangle_a|0\rangle_b$. 
These states thus can be generated by injecting a Fock state $|N\rangle$ and a vacuum state $|0\rangle$ as inputs 
into a beam splitter with the transmittance $\cos^2{r}=(1+\lambda)/2$. 
Since the photon number difference is obtained as $\langle N_-\rangle=\lambda N$ from Eq.~(\ref{eqn:N-diff}), 
it is easily checked that the states $|\Psi\rangle_{N,N}$ all violate both the inequalities~(\ref{eqn:Hillery}) 
and~(\ref{eqn:opt}).  

The states $|\Psi\rangle_{N,N}$ actually correspond to the SU(2)-coherent states considered in Ref.~\cite{Gerry}. 
For $N=1$, the state $|\Psi\rangle_{1,1}$ is an arbitrary superposition of the single-photon states,  
$|\Psi\rangle_{1,1}=\cos{r}|1\rangle_a|0\rangle_b+\sin{r}|0\rangle_a|1\rangle_b$\cite{Lee}.\\ 

(b) case of $m=N-1$\\
In this case, we find 
\begin{eqnarray}
|\Psi'\rangle_{N,N-1}=\frac{1}{\sqrt{\alpha^2+\beta^2}}
\left(\alpha|N-1,1\rangle+\beta|N,0\rangle\right),
\end{eqnarray}
where $\alpha=\sqrt{1-\lambda^2}$, $\beta=-\lambda \sqrt{N}$, 
and  
\begin{eqnarray}
\langle N_-\rangle=\frac{\lambda\left[3N-2+(N^2-3N+2)\lambda^2\right]}{1+(N-1)\lambda^2}. 
\end{eqnarray}
Now, the inequality~(\ref{eqn:Hillery}) is violated only for the squeezing parameter 
$\lambda>\lambda_c=\frac{1}{\sqrt{N-1}}$. 
On the other hand, the optimal inequality~(\ref{eqn:opt}) is violated for any values of $\lambda$, 
showing that the inequality~(\ref{eqn:opt}) is stronger than~(\ref{eqn:Hillery}). 

Instead of seeking further analytic expressions, 
we numerically study the violation of the inequalities by the SU(2) minimum-uncertainty states, 
and the results are displayed in Figs.~1-3. 
We calculate the quantities $Q$ and $R$ defined by 
\begin{eqnarray}
Q&\equiv&\frac{(\Delta J_x)^2+(\Delta J_y)^2}{\langle N_+\rangle/2}-1,\nonumber\\ 
R&\equiv&\frac{(1+4(\Delta J_x)^2)(1+4(\Delta J_y)^2)}{(1+\langle N_+\rangle)^2}-1.
\label{eqn:QR}
\end{eqnarray}
The negativity of $Q$ and $R$ represents the violation of the inequality~(\ref{eqn:Hillery}) and~(\ref{eqn:opt}), respectively.

\begin{figure}
\includegraphics*[width=3.3in,keepaspectratio=true]{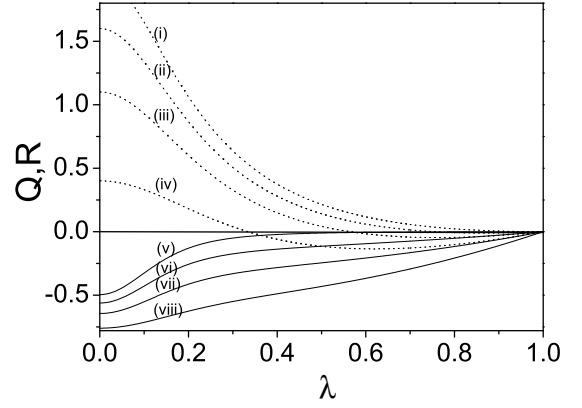}
\caption{(i)-(iv) (dotted lines) $Q$- and (v)-(viii) (solid lines) $R$-values, defined in Eq.~(\ref{eqn:QR}), 
as functions of the squeezing parameter $\lambda$ for the states $|\Psi\rangle_{N=10,m}$: 
(i),(v) $m=5$, (ii),(vi) $m=3$, (iii),(vii) $m=2$, and (iv),(viii) $m=1$. }
\label{fig:desu1}
\end{figure}

In Fig.~\ref{fig:desu1}, we plot $Q$ and $R$ for $|\Psi\rangle_{N=10,m}$ 
as functions of the squeezing parameter $\lambda$. 
We have found that $Q$ and $R$ values for $|\Psi\rangle_{N,m}$ are the same 
as the corresponding ones for $|\Psi\rangle_{N,N-m}$ in general. 
For example, the curves (ii) and (vi) in Fig.~\ref{fig:desu1} represent $Q$ and $R$ for the states 
$|\Psi\rangle_{N=10,3}$ and $|\Psi\rangle_{N=10,7}$ alike. 

\begin{figure}
\includegraphics*[width=3.3in,keepaspectratio=true]{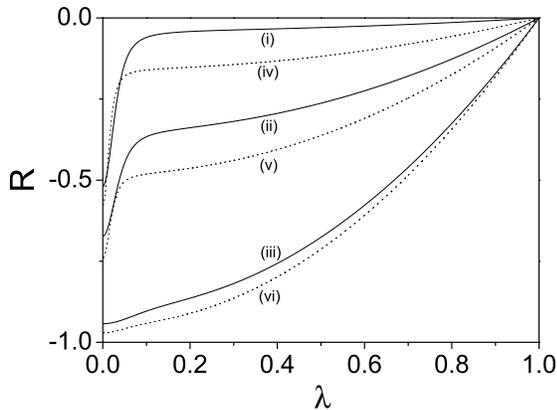}
\caption{The $R$ values, defined in Eq.~(\ref{eqn:QR}), 
as a function of the squeezing parameter $\lambda$. Solid (dotted) curves for the states $|\Psi\rangle_{N=50,m}$ 
($|\Psi\rangle_{N=100,m}$): 
(i)$m=20,30$, (ii) $m=10,40$, (iii) $m=1,49$, and (iv) $m=30,70$, (v) $m=15,85$, (vi) $m=1,99$. }
\label{fig:desu2}
\end{figure}

We can show that the $R$ values are always negative, indicating that the inequality~(\ref{eqn:opt}) is violated, 
for arbitrary states $|\Psi\rangle_{N,m}$ with any degree of squeezing. 
In Fig.~\ref{fig:desu2}, for instance, we plot the R values for the states $|\Psi\rangle_{N=50,m}$ 
and $|\Psi\rangle_{N=100,m}$. 
As the degree of squeezing is increased, i.e., $\lambda\rightarrow0$, the negativity of $R$ values deepens. 
In the extreme squeezing of $\lambda=0$ ($(\Delta J_x)^2=0$), we obtain $|\Psi'\rangle_{N,m}=|m\rangle_a|N-m\rangle_b$ 
from Eqs.~(\ref{eqn:eigenstate}) and~(\ref{eqn:coeff}). 
Then, by a direct calculation of $(\Delta J_y)^2$, the extreme $R$ value is given by 
$R=-\frac{N^2+N(1-2m)+2m^2}{(1+N)^2}$. 

\begin{figure}
\includegraphics*[width=3.3in,keepaspectratio=true]{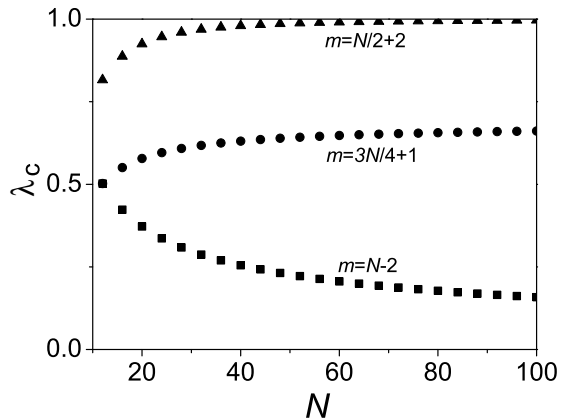}
\caption{The critical value $\lambda_c$ as a function of total photon number $N(\ge12)$  
with $m=N/2+2$ (triangle), $m=3N/4+1$ (circle), and $m=N-2$ (square), for the states $|\Psi\rangle_{N,m}$. 
The inequality~(\ref{eqn:Hillery}) is violated for the squeezing parameter $\lambda>\lambda_c$. }
\label{fig:desu3}
\end{figure}

On the other hand, the $Q$ values become negative only for a certain range of $\lambda>\lambda_c$, 
where the critical value $\lambda_c$ satisfies the equality in Eq.~(\ref{eqn:Hillery}). 
Arranging terms, we find that $\lambda_c$ satisfies the equation 
\begin{eqnarray}
(\tilde m+3m+4\Lambda^2 m){_2F_1}\left[m+1,-\tilde m,1,-\Lambda^2\right]\nonumber\\
=2m(1+2\Lambda^2)(\tilde m+1){_2F_1}\left[m+1,-\tilde m,2,-\Lambda^2\right], 
\end{eqnarray}
where $\tilde m\equiv N-m$ and $\Lambda=-\frac{\lambda}{\sqrt{1-\lambda^2}}$.
For fixed $N$, $\lambda_c$ varies with the difference $|N/2-m|$, 
and more precisely, $\lambda_c$ decreases with the increasing value of $|N/2-m|$. 
In other words, the range of the squeezing parameter $\lambda$, 
for which the inequality~(\ref{eqn:Hillery}) is violated, 
becomes broader for a larger difference $|N/2-m|$. 
In Fig.~\ref{fig:desu3}, we plot the critical value $\lambda_c$ 
for different $\{N,m\}$-values. 

In conclusion, we have shown that the optimal inequality~(\ref{eqn:opt}) can detect entanglement 
for any arbitrary SU(2) minimum-uncertainty states, 
whereas the inequality~(\ref{eqn:Hillery}) has some limited applicability of detecting entanglement.
\\

\section{Measurement scheme}

In this section, we discuss the generation of the SU(2) squeezed states and 
the measurement of the observables $J_x$ and $J_y$ necessary for testing the criterion~(\ref{eqn:opt}).  
The experimental schemes for generating the SU(2) squeezed states have been suggested 
for atomic system\cite{Agarwal2} and for optical fields\cite{Luis}. 
In particular, Luis and Perina proposed to generate the SU(2) coherent states, 
$|\Psi\rangle_{N,N}$ in our notation, using two parametric down converters 
(PDCs) with the aligned idler modes\cite{Luis}. In their scheme, however, 
it is required to project the idler mode to a photon number state, 
which seems hard to implement in practice. 
They also showed that other SU(2) squeezed states can be produced in the signal modes 
within the same setup, when a beam splitter is inserted between the PDCs and 
projective measurements are performed on the two idler modes.

Another method to generate the SU(2) coherent states is to inject a photon number state 
$|N,0\rangle$ to a beam splitter, as addressed in the subsection IV.~B.~(ii)-(a). 
It is, however, also demanding to produce the photonic Fock states. 
Recently, an experimental scheme was proposed to extract a Fock state from an input coherent state 
using linear optics and projective measurements\cite{Sanaka}. 

Let us now discuss how to measure the observables $J_x=\left(a^\dag b+ab^\dag\right)/2$ and 
$J_y=\left(a^\dag b-ab^\dag\right)/2i$ in experiment. 
M.~Hillery showed that the variances $\Delta J_x$ and $\Delta J_y$ can be measured 
via a nonlinear interaction, i.e., the difference-frequency generation described by the Hamiltonian 
$H=\hbar g(a^\dag bc^\dag+ab^\dag c)$ \cite{Hillery3}. 
Specifically, the variances of the two orthogonal quadrature amplitudes for mode $c$, 
which can be measured in homodyne detection, correspond to $\Delta J_x$ and $\Delta J_y$, respectively.   

\begin{figure}
\includegraphics*[width=3in,keepaspectratio=true]{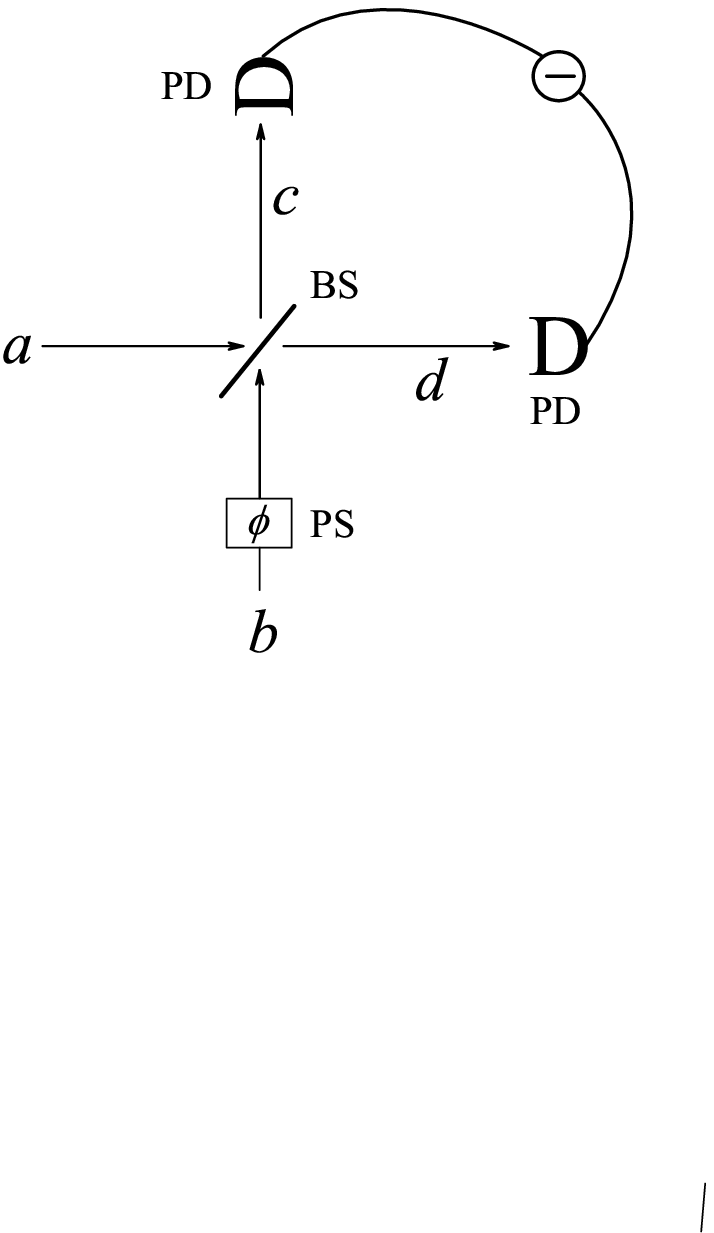}
\vspace{-2.3in}
\caption{Experimental scheme for measuring $J_x=\left(a^\dag b+ab^\dag\right)/2$ and 
$J_y=\left(a^\dag b-ab^\dag\right)/2i$. BS: beam-splitter, PS: phase-shifter. $J_x$ ($J_y$) can be detected, 
with the phase shift $\phi=0$ ($\phi=\frac{\pi}{2}$), 
by measuring the photon-number difference $c^\dag c-d^\dag d$ at the output.}
\label{fig:desu4}
\end{figure}

Alternatively, we can use linear optical devices along with photon detectors 
[Fig.~\ref{fig:desu4}]. 
The mode $b$ first goes through a phase shifter and 
the two modes $a$ and $b$ are injected to a 50:50 beam splitter. 
The output modes $c$ and $d$ are given by $c=\frac{1}{\sqrt{2}}(a+be^{-i\phi})$ and 
$d=\frac{1}{\sqrt{2}}(-a+be^{-i\phi})$. 
One then measures the photon number difference at the output, i.e., 
$c^\dag c-d^\dag d=a^\dag be^{-i\phi}+ab^\dag e^{i\phi}$, 
which becomes $2J_x$ ($2J_y$) for $\phi=0$ ($\phi=\frac{\pi}{2}$). 
In fact, this scheme is none other than the typical homodyne detection 
when one of the input mode, say $b$, is replaced by a large intensity coherent field.  
Although this method does not require the nonlinear interaction as compared with Hillery's scheme, 
photon counting is usually less efficient than homodyne detection. 
The photon detectors currently available in the laboratory are known to be sensitive only to low photon numbers. 
So the measurement via linear optics seems to be rather demanding for the SU(2) squeezed states 
with large photon numbers. 
We, however, believe that the experimental progress in this direction will be continuously made\cite{Mukunda}. 

\section{Summary} 

In this paper, we have derived a class of inequalities~(\ref{eqn:uncertainty5}), 
from the SU(1,1) and the SU(2) algebra in conjunction with the partial transposition, 
that all separable CV states must satisfy. 
The strongest inseparability criterion among them is the same 
as the one derived by Agarwal and Biswas\cite{Agarwal}. 
We have shown that this optimal condition can detect entanglement for a broad class of non-Gaussian entangled states, 
that is, the SU(2) minimum-uncertainty states with any degree of squeezing. 
Examples of such states include the ones produced by superposing a Fock state $|N\rangle$ and a vacuum state $|0\rangle$ at a beam splitter.  
For comparison, we also considered the inequality~(\ref{eqn:Hillery}), also independently derived by Hillery and Zubairy, 
which is found to be capable of detecting inseparability only for a certain range of the squeezing parameter. 
We have proposed a linear optical scheme with photon detectors 
to test the optimal criterion in experiment.

We acknowledge the financial support from the Korea Ministry of Science and Technology under the contract NC33520.\\

*email:phylove00@gmail.com


\begin{references}
\bibitem{nnha}Note that the squeezing in this paper is referred only in the two-mode sense of 
Eq.~(\ref{eqn:eigen-rel7}) for SU(2) algebra, and should be distinguishied from the usual quadrature squeezing.
\bibitem{nha1} In fact, the superposition of the states $|\Psi'\rangle_{N,m}$ for different $\{N,m\}$s 
but with the same eigenvalue $\beta$ ( i.e., the same $2m-N$), e.g., 
$|\Psi'\rangle=\alpha_1|\Psi'\rangle_{2,0}+\alpha_2|\Psi'\rangle_{4,1}$, 
also becomes an eigenstate that satisfies Eq.~(\ref{eqn:eigen-newnew}). 
In this paper, however, we will not deal with those superposition states.
\end{references}
\end{document}